\documentclass[floatfix,aps,amsmath,nofootinbib,twocolumn,10pt]{revtex4}
\pdfoutput=1
\usepackage{listings}
\usepackage{graphicx}
\usepackage{bm}
\usepackage{rotating}
\usepackage{array}
\usepackage{amsmath}
\usepackage{amssymb} 
\usepackage{mathrsfs} 
\usepackage{cancel}
\usepackage{color}

\def\({\left(}
\def\){\right)}
\def\[{\left[}
\def\]{\right]}

\def\e{\begin{equation}}
\def\q{\end{equation}}
\def\m{\begin{eqnarray}}
\def\n{\end{eqnarray}}

\begin{document}


\title{Implications for cosmology from Ground-based Cosmic Microwave Background observations}

\author{ Ke Wang$^{1}$\footnote{wangkey@lzu.edu.cn} and Qing-Guo Huang$^{2,3,4,5}$ \footnote{huangqg@itp.ac.cn}}
\affiliation{$^1$ Institute of Theoretical Physics \& Research Center of Gravitation,\\ Lanzhou University, Lanzhou 730000, China\\
$^2$ CAS Key Laboratory of Theoretical Physics,\\ Institute of Theoretical Physics, \\Chinese Academy of Sciences, Beijing 100190, China\\
$^3$ School of Physical Sciences, \\University of Chinese Academy of Sciences,\\ No. 19A Yuquan Road, Beijing 100049, China\\
$^4$ Synergetic Innovation Center for Quantum Effects and Applications, Hunan Normal University, 36 Lushan Lu, 410081, Changsha, China\\
$^5$ Center for Gravitation and cosmology, College of Physical Science and Technology, Yangzhou University, 88 South University Ave., 225009, Yangzhou, China\\
}

\date{\today}

\begin{abstract}

Cosmic Microwave Background (CMB) anisotropy encodes a lot of information about our Universe. In this paper we take the ground-based CMB observations (GCMB), including the South Pole Telescope (SPT), SPTpol and the Atacama Cosmology Telescope Polarimeter (ACTPol), as a new probe to the CMB anisotropy independent of two satellite observations, i.e. Wilkinson Microwave Anisotropy Probe (WMAP) and Planck. The combination of current GCMB data is consistent with WMAP and Planck. In the spatially flat $\Lambda$CDM model, the Hubble constant is $H_0=69.72\pm 1.63$ km/s/Mpc at $68\%$ confidence level (CL). Combining with baryon acoustic oscillation (BAO) and the Pantheon sample of Type Ia supernovae (SN), we find that $H_0=68.40\pm 0.58$ km/s/Mpc ($68\%$ CL) in the spatially flat $\Lambda$CDM cosmology which has a tension with local measurement given by Riess et al. in 2019 at $3.7\sigma$ level, and $\Omega_k=-0.0013\pm 0.0039$ and $N_{\rm{eff}}=2.90\pm 0.41$ ($68\%$ CL) in the extended cosmological models. 

\end{abstract}

\pacs{???}

\maketitle


\section{Introduction}
\label{introduction}
The free streaming of photons from the last scattering surface preserves the acoustic oscillations of the photon-baryon fluid in the early universe, which results in the temperature anisotropy of cosmic microwave background (CMB). Moreover, the quadrupole radiation with large and small wavenumber can be polarized at recombination and reionization epoch respectively, which leads to the polarization anisotropy of CMB. 
Therefore, measurements of the temperature and polarization anisotropy of CMB provide the information about the primordial perturbations, the ionization history, the composition and evolution of the Universe and its geometry. 

So far, two CMB anisotropy final data releases from Wilkinson Microwave Anisotropy Probe (WMAP) satellite \cite{Bennett:2012zja} and Planck satellite \cite{Aghanim:2018eyx} respectively have confirmed the standard spatially-flat six-parameter $\Lambda$CDM cosmology with a power-law spectrum of adiabatic scalar perturbations. 
Here the ground-based CMB observations (GCMB), including the South Pole Telescope (SPT) \cite{Story:2012wx}, SPTpol \cite{Henning:2017nuy} and the Atacama Cosmology Telescope Polarimeter (ACTPol) \cite{Louis:2016ahn}, is taken as a new probe to the CMB anisotropy approximately independent of WMAP and Planck. 
More precisely, GCMB includes the CMB temperature anisotropy power spectrum (TT) over the multipole range $650<\ell<3000$ from the 2500-square-degree SPT-SZ survey \cite{Story:2012wx}, the E-mode polarization angular auto-power spectrum (EE) and temperature-E-mode cross-power spectrum (TE) of CMB over the multipole range $50<\ell<8000$ from the 500-square-degree SPTpol data \cite{Henning:2017nuy}, and the two-season temperature and polarization angular power spectra over the multipole range $350<\ell<4125$ measured by ACTPol from 548-square-degree of sky  \cite{Louis:2016ahn}.
Since SPT-SZ survey covered a $\sim2500~\rm {deg}^2$ region of sky between declinations (dec) of $-65^{\circ}$ and $-40^{\circ}$ and right ascensions (RA) of $20~\rm h$ and $7~\rm h$, SPTpol survey field is a $500~\rm {deg}^2$ patch of sky spanning $4~\rm h$ of RA, from $22~\rm h$ to $2~\rm h$, and $15^{\circ}$ of dec, from $-65^{\circ}$ and $-50^{\circ}$, ACTPol covered $548~\rm {deg}^2$ with coordinates $-7.2^{\circ}<\rm{dec}<4^{\circ}$ and $23~\rm h< \rm{RA}<3~\rm h$ and BICEP2/Keck Array CMB polarization experiments \cite{Ade:2018gkx} covered a $\sim400~\rm {deg}^2$ region of sky centered at RA $0~\rm h$ and dec $-57.5^{\circ}$, there is an overlap between SPT and BICEP2/Keck Array CMB polarization experiments in sky coverage. Therefore we exclude data from BICEP2/Keck Array, but there is almost no correlation between SPT-SZ (or SPTpol) and ACTPol for TT (or EE) spectrum in GCMB data. 


Furthermore, in order to break the degeneracies among the cosmological parameters, we will also combine the baryon acoustic oscillation (BAO) data and the Pantheon sample of Type Ia supernovae (SN) data. The BAO data includes the SDSS DR7 Main Galaxy Sample (MGS) \cite{Ross:2014qpa}, the Six-degree-Filed Galaxy Survey (6dFGS) \cite{Beutler:2011hx}, the anisotropic BAO analysis from Baryon Oscillation Spectroscopic Survey (BOSS) data  release 12 (DR12) \cite{Alam:2016hwk}, the correlations of quasar sample in extended Baryon Oscillation Spectroscopic Survey (eBOSS) data release 14 (DR14) \cite{Hou:2018yny} and the correlations of Ly$\alpha$ absorption in eBOSS DR14 \cite{Agathe:2019vsu}. The `Pantheon Sample' consisting of a total of 1048 SN Ia ranging from $0.01<z<2.3$, which can be separated into five  subsamples: PS1 with 279 SN Ia ($0.03<z<0.068$) \cite{Scolnic:2017caz}, SDSS with 335 SN Ia ($0.04 < z < 0.42$) \cite{Frieman:2007mr,Kessler:2009ys}, SNLS with 236 SN Ia ($ 0.08< z <1.06$) \cite{Conley:2011ku,Sullivan:2011kv}, Low-z with 172 SN Ia ($z<0.08$) \cite{Riess:1998dv,Jha:2005jg,Hicken:2009dk,Hicken:2009df,Hicken:2012zr,Contreras:2009nt,Folatelli:2009nm,Stritzinger:2011qd} and HST with 26 SN Ia ($z>1$) \cite{Suzuki:2011hu,Riess:2006fw,Rodney:2014twa,Graur:2013msa,Riess:2017lxs}.

In this paper we will constrain the cosmological parameters from GCMB and combining with BAO and SN data by using the 2019 July version of the Markov Chain Monte Carlo (MCMC) package CosmoMC \cite{Lewis:2002ah}. The following part is organized as follows. In Sec.~\ref{six}, the base $\Lambda$CDM model constraints from the combination of ground-based experiments are presented. In Sec.\ref{seven}, we consider two one-parameter extensions to the base $\Lambda$CDM model and then their parameter constraints are given. A brief summary is given in Sec.\ref{sum}. 

\section{Constraints on the base $\Lambda$CDM cosmology}
\label{six}

\begin{figure*}[]
\begin{center}
\includegraphics[height=15cm]{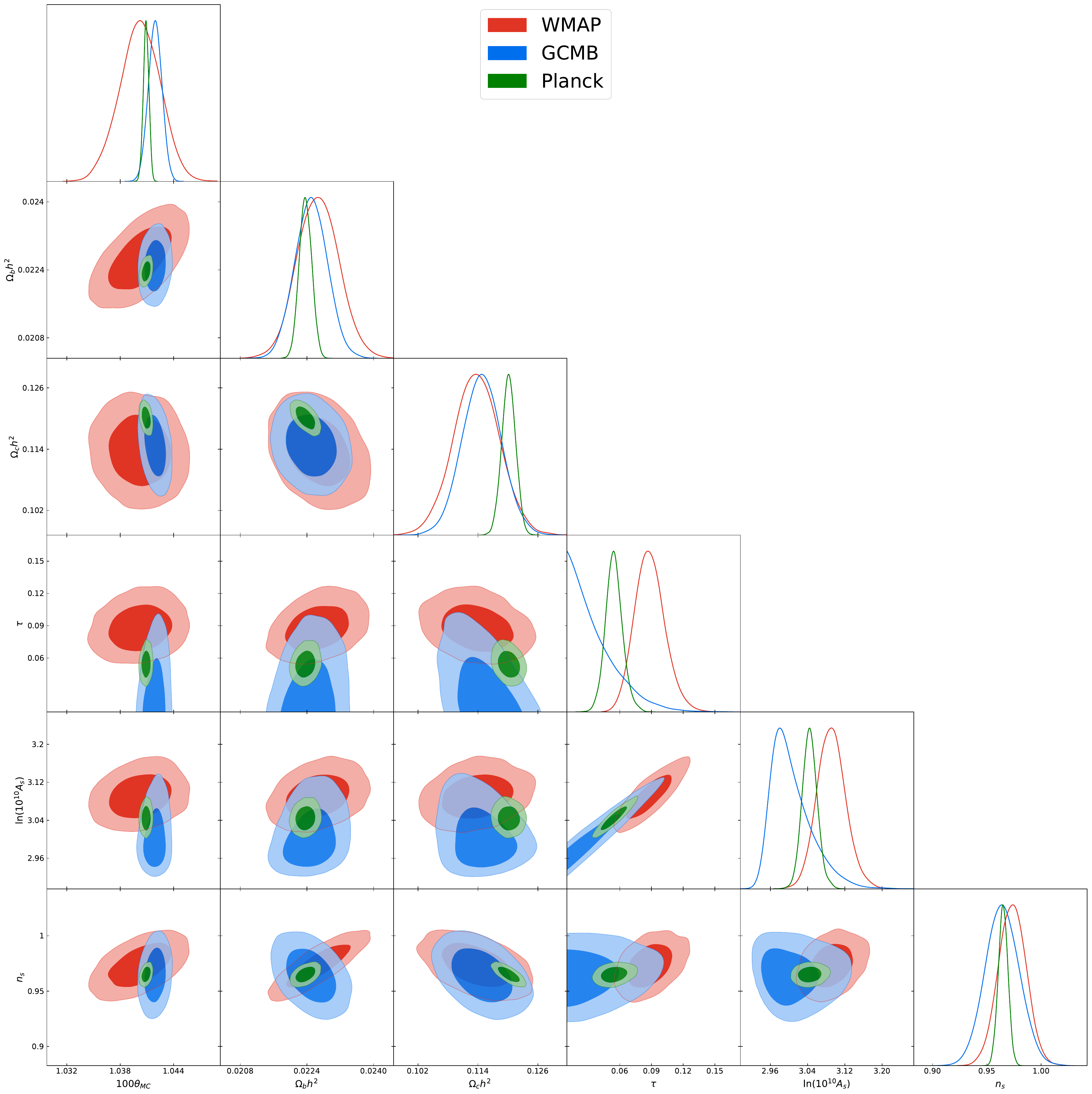}
\end{center}
\caption{Comparison of the base $\Lambda$CDM model parameter constraints from GCMB (blue), WMAP (red) and Planck (green).}
\label{fig:3CMB}
\end{figure*}

First of all, we focus on the standard spatially-flat six-parameter $\Lambda$CDM cosmology which is also denoted by the base $\Lambda$CDM in literature. The six parameters in this model are 
\m
\{\Omega_b h^2,\ \Omega_c h^2,\ 100\theta_{\rm MC},\ \tau,\ \ln(10^{10}A_s),\ n_s \}, 
\n
where $\Omega_b h^2$ is the physical density of baryons today, $\Omega_c h^2$ is the physical density of cold dark matter today, $\theta_{\rm MC}$ is the ratio between the sound horizon and the angular diameter distance at last scattering, $\tau$ is the Thomson scatter optical depth due to reionization, $A_s$ is the amplitude of the power spectrum of primordial curvature perturbations at the pivot scale $k_0= 0.05\ \rm Mpc^{-1}$, and $n_s$ is the spectral index of the scalar fluctuations. Assuming flat priors for all of these parameters and setting the R-1 convergence value as $0.01$, we adopt MCMC to work out the parameter estimations which are summarized in Tab.~\ref{tab:3CMB}.
\begin{table*}[!htp]
\centering
\renewcommand{\arraystretch}{1.5}
\begin{tabular}{c||c|c||c|c}
\hline\hline
  Parameter & Priors for GCMB & GCMB & WMAP  & Planck \\
\hline
$\Omega_b h^2$&[0.005, 0.1]&$0.02250\pm0.00038$                                           & $0.02264\pm0.00050$&$0.02236\pm0.00015$ \\
$\Omega_c h^2$&[0.001, 0.99]&$0.1148\pm0.0039$                                & 
$0.1138\pm0.0045$ &   $0.1202\pm0.0014$ \\
$100\theta_{\rm MC}$&[0.5, 10]&$1.04193\pm0.00075$                              & $1.04023\pm0.00222$ &  $1.04090\pm0.00031$\\
$\ln(10^{10}A_s)$&[1.61, 3.91]&$3.007^{+0.024}_{-0.053}$                                        & $3.091\pm0.031$  &    $3.045\pm0.016$\\
$n_s$&[0.8, 1.2]&$0.9643\pm0.0154$                                        & 
$0.9734\pm0.0124$ &    $0.9649\pm0.0044$\\
$\tau$&[0.01, 0.8]&$<0.0807~(95\%)$                                        & 
$0.0885\pm0.0141$  &    $0.0544^{+0.0070}_{-0.0081}$\\
\hline
\hline
\end{tabular}
\caption{$\Lambda$CDM model parameter constraints at $68\%$ CL from GCMB, WMAP and Planck.} 
\label{tab:3CMB}
\end{table*}
Here we also list the results from WMAP and Planck. The contour plots for the cosmological parameters in the base $\Lambda$CDM cosmology are given in Fig.~\ref{fig:3CMB}. 

From both Tab.~\ref{tab:3CMB} and Fig.~\ref{fig:3CMB}, roughly speaking, the GCMB is consistent with both WMAP and Planck, and the precision of current GCMB data is comparable with WMAP, but worse than Planck. 


\subsection{Hubble constant}

\begin{figure}[]
\begin{center}
\includegraphics[height=8 cm]{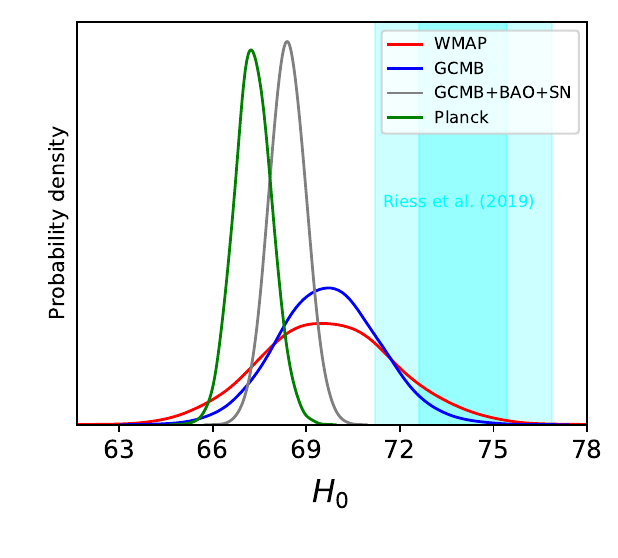}
\end{center}
\caption{Constraints on $H_0$ as a derived parameter of the base-$\Lambda$CDM model, from WMAP (red), GCMB (blue), GCMB+BAO+SN (grey), and Planck (green). And the cyan bands correspond to the local Hubble constant measurement from \cite{Riess:2019cxk}.}
\label{fig:5H0}
\end{figure}

The Hubble constant $H_0$ denotes the expansion rate of the Universe at present. In 2018, Planck final data release \cite{Aghanim:2018eyx} implies 
\m
H_0=67.27\pm 0.60\ \hbox{km/s/Mpc} 
\n
in spatially-flat $\Lambda$CDM cosmology at $68\%$ confidence level (CL) which has an around $4.4\sigma$ tension with local measurement by Riess et al. in \cite{Riess:2019cxk}, namely 
\m
H_0=74.03\pm1.42\ \hbox{km/s/Mpc} 
\n
at $68\%$ CL. In \cite{Addison:2015wyg}, the authors found that $H_0=69.7\pm 1.7$ km/s/Mpc from $\ell<1000$ Planck data and $H_0=64.1\pm 1.7$ km/s/Mpc from $\ell\geq1000$ Planck data. It implies that the tension between Planck data and local measurement may mainly come from the Planck data at small scales. And the measurement on $H_0$ from large-scale data $\ell<1000$ of Planck is nicely consistent with WMAP data, i.e. $H_0=70.7\pm 2.2$ km/s/Mpc. Here, GCMB including high-$\ell$ CMB data provides an independent CMB measurement on small scales, and we find 
\begin{equation}
H_0=(69.72\pm 1.63)\ \hbox{km/s/Mpc},
\end{equation}
at $68\%$ CL, which is consistent with both WMAP and $\ell<1000$ Planck data. See Fig.~\ref{fig:5H0}. 

\begin{figure*}[]
\begin{center}
\includegraphics[height=4.2cm]{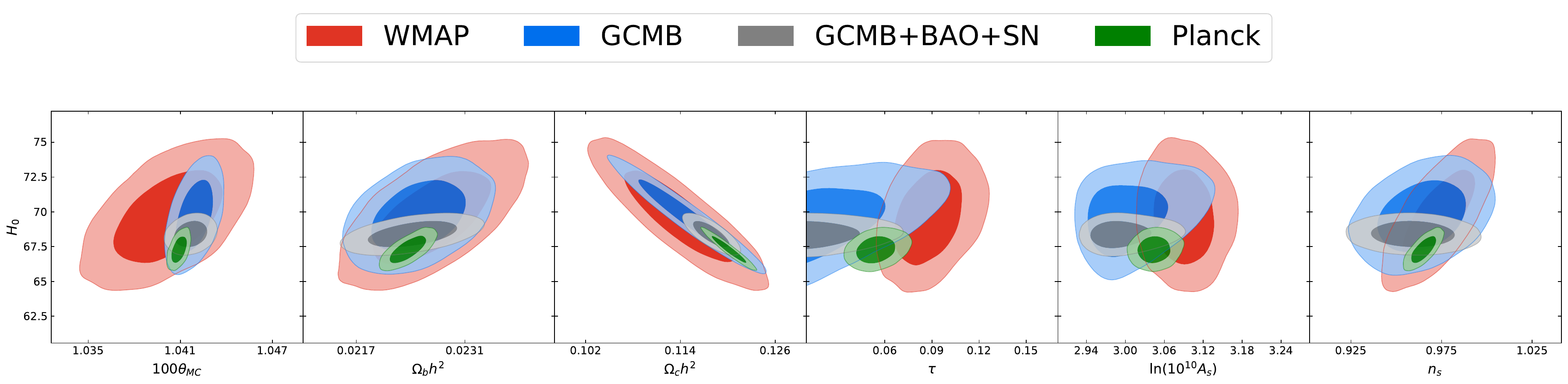}
\end{center}
\caption{Comparison of the base $\Lambda$CDM model parameter constraints from WMAP (red), GCMB (blue), GCMB+BAO+SN (grey), and Planck (green). }
\label{fig:H0}
\end{figure*}

Furthermore, in order to break the degeneracy among the cosmological parameters, we take some low-redshift data, such as BAO and SN, into account. 
Combining GCMB, BAO and SN data, we obtain a $0.9\%$ constraint, i.e. 
\begin{equation}
H_0=(68.40\pm 0.58)\ \hbox{km/s/Mpc}
\end{equation}
at $68\%$ CL. See Fig.~\ref{fig:H0} in detail. 
It indicates that the determination of $H_0$ from GCMB+BAO+SN is consistent with Planck \cite{Aghanim:2018eyx}, WMAP+BAO \cite{Zhang:2018air} and low-redshift data only \cite{Zhang:2018jfu,Zhang:2019cww}. Our result imply that there is a strong tension on the Hubble constant between all of CMB data and local measurement. In a word, this tension may come from some unknown systematic errors, or the new physics beyond the standard cosmology \cite{Qing-Guo:2016ykt,Liu:2019awo,Kumar:2016zpg,Vagnozzi:2017ovm,Lin:2017bhs,DiValentino:2017rcr,Guo:2018uic,Vagnozzi:2018jhn,Feeney:2018mkj,DiValentino:2018jbh,Vagnozzi:2019ezj,Ghosh:2019tab,DiValentino:2019ffd,Escudero:2019gvw,DiValentino:2019jae} etc.

\subsection{Reionization optical depth}

\begin{figure}[]
\begin{center}
\includegraphics[height=7.5cm]{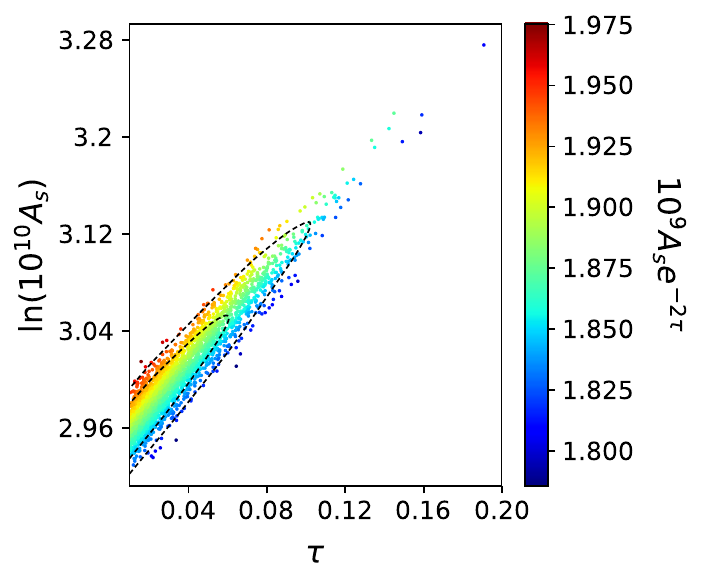}
\end{center}
\caption{Constraints on $\tau$, $\ln(10^{10}A_s)$ and $10^9A_se^{-2\tau}$ in the base $\Lambda$CDM model from GCMB.}
\label{fig:tau}
\end{figure}

Since the average observed CMB power spectrum amplitude scales with the parameter combination $A_s e^{-2\tau}$, there is a strong degeneracy between $A_s$ and $\tau$. It is also the case for GCMB, as shown in Fig.~\ref{fig:tau}. The reionization occurred at around $z\sim 10$, and then the large-scale anisotropies in polarization are sensitive to $\tau$. Unfortunately, in GCMB data, the E-mode polarization spectrum of SPTpol over the multipole range $50<\ell<8000$ are not so useful to significantly constrain the optical depth $\tau$. Therefore, GCMB itself cannot provide a good constraint on the optical depth, namely 
\m
\tau<0.0807 
\n
at $95\%$ CL, which is still consistent with the optical depth constrained by the large-scale polarization measurements from final Planck release. 

\subsection{Anti-correlation between $n_s$ and $\Omega_bh^2$}

From Tab.~\ref{tab:3CMB} and Fig.~\ref{fig:3CMB}, we see that GCMB prefers a red-tilted scalar power spectrum whose spectral index is 
\m
n_s=0.9643\pm 0.0154
\n
at $68\%$ CL. 
It implies that the scalar power spectrum should be suppressed with $\ell$ increasing. Due to the most distinctive imprints on spectra on scales $\ell<1000$ leaved by $\Omega_b h^2$ which affect the relative heights of odd peaks and even peaks, there is a mild correlation between $n_s$ and $\Omega_b h^2$ for WMAP and Planck. However, there is a mild anti-correlation, instead of correlation, between $n_s$ and $\Omega_b h^2$ for GCMB as shown in Fig.~\ref{fig:3CMB}.
The reason is that the spectrum of GCMB covers a much higher multipole range than that of WMAP and Planck, and a higher value of $\Omega_b h^2$ enhances the spectra on very small scales by reducing the diffusion length (or increasing the damping wavenumber), which can be compensated by a redder-tilted scalar power spectrum.

\section{Extensions to the base $\Lambda$CDM model}
\label{seven}

Even though the spatially-flat six-parameter standard cosmology is consistent with GCMB, it is still worthy exploring whether there are some clues for new physics in the data. Here we consider two one-parameter extensions to the base $\Lambda$CDM model, i.e. the spatial curvature energy density $\Omega_k$ and the effective number of relativistic degree of freedom $N_{\rm eff}$.

\subsection{Spatial curvature}

How to explain the flatness of our Universe is one of the crucial motivations for inflationary cosmology \cite{Guth:1980zm,Starobinsky:1979ty,Linde:1981mu} (see \cite{Li:2019efi,Li:2019vlb} for some recent investigations). In general, inflationary models have a large number of e-folds, and hence our Universe should be very closed to spatially flat. 

\begin{figure}[]
\begin{center}
\includegraphics[height=7.5cm]{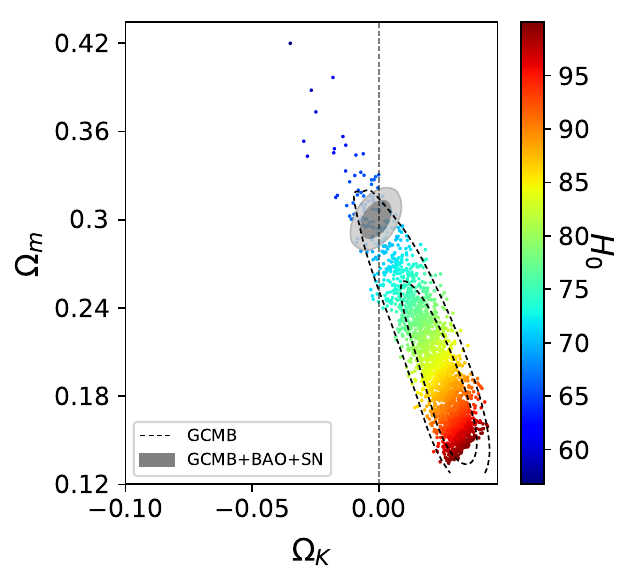}
\end{center}
\caption{Constraints on the extension to the base $\Lambda$CDM model with an additional parameter $\Omega_K$ from GCMB (the color points and the dashed contours) and GCMB+BAO+SN (grey).}
\label{fig:k}
\end{figure}

Due to the geometric degeneracy, CMB data only cannot constrain the spatial curvature $\Omega_k$ and the Hubble constant $H_0$ well. 
From GCMB data only, we find a slight preference for an open Universe which drives the value of $H_0$ towards a larger value, namely
\m
\Omega_k&=&0.0218\pm 0.0107, \\
H_0&=&(85.12\pm 8.83)\ \hbox{km/s/Mpc},
\n
at $68\%$ CL. See Fig.~\ref{fig:k} as well. 
On the other hand, it is well-known that the addition of probes of late time physics can break the geometric degeneracy effectively. Here we gives the constraints on $\Omega_k$ and $H_0$ from GMCB+BAO+SN dataset as follows 
\m
\Omega_k&=&-0.0013\pm 0.0039,\\
H_0&=&(68.25\pm 0.75)\ \hbox{km/s/Mpc},
\n
at $68\%$ CL. See the grey contours in Fig.~\ref{fig:k}. We see that a spatially-flat Universe is preferred at high statistical CL once the BAO and SN datasets are combined.

\subsection{Effective number of relativistic species}

The total energy density of radiation in the Universe is 
\m
\rho_{\rm{rad}}=\[1+{7\over 8}\({4\over 11}\)^{4\over 3} N_{\rm{eff}}\]\rho_\gamma, 
\n
which is a sum of the CMB photon energy density $\rho_\gamma$ and the energy density of standard model (SM) neutrinos if $N_{\rm{eff}}=3.046$. Here the neutrino mass are considered to be small \cite{Huang:2015wrx,Xu:2016ddc}. If $N_{\rm{eff}}>3.046$, it may imply the existence of extra relativistic degree of freedom.

\begin{figure}[]
\begin{center}
\includegraphics[height=7.5cm]{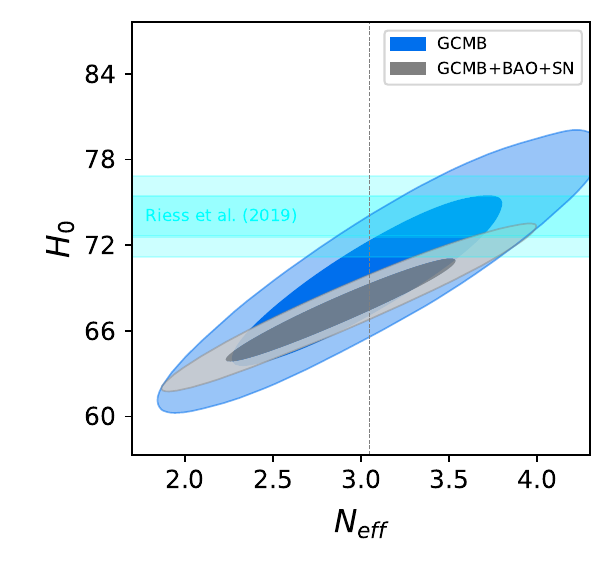}
\end{center}
\caption{Constraints on the extension to the base $\Lambda$CDM model with an additional parameter $N_{\rm eff}$ from  GCMB (the blue) and GCMB+BAO+SN (grey). The cyan bands
show the local Hubble parameter measurement $H_0=(74.03\pm1.42)$ km/s/Mpc from \cite{Riess:2019cxk}. }
\label{fig:nnu}
\end{figure}

In this subsection, we explore the constraints on $N_{\rm{eff}}$ from GCMB. Our results are illustrated in Fig.~\ref{fig:nnu}. 
For $N_{\rm{eff}}>3.046$, GCMB data prefer higher values of $H_0$ because higher value of $N_{\rm{eff}}$ yields a smaller sound horizon at recombination and the Hubble constant need rise to keep the acoustic scale fixed at the observed value. More precisely, the constraints on $N_{\rm{eff}}$ and $H_0$ are 
\m
N_{\rm eff}&=&3.08\pm 0.49,\\
H_0&=&(69.95\pm 3.80)\ \hbox{km/s/Mpc},
\n
at $68\%$ CL from GCMB data, and 
\m
N_{\rm eff}&=&2.90\pm 0.41,\\
H_0&=&(67.60\pm 2.26)\ \hbox{km/s/Mpc}, 
\n
at $68\%$ CL by combing GCMB data with BAO and SN data. Our results imply that more (or less) relativistic degree of freedom may relax (or aggravate) the tension on the Hubble constant between the local measurement and the global fitting from CMB.

\section{Summary}
\label{sum}

In this paper, we constrain the cosmological parameters in the six-parameter spatially-flat $\Lambda$CDM cosmology from the current GCMB data. Compared to the results from WMAP and Planck, we find that these three CMB observations are consistent with each other. In particular, CMB datasets systematically prefer a lower value of the Hubble constant in the base $\Lambda$CDM cosmology compared to the local measurement \cite{Riess:2019cxk}. 
Moreover, due to the lack of polarization anisotropy data on the very large scales, there is a strong degeneracy between $A_s$ and $\tau$ for GCMB, and due to the complete ``damping tail" of CMB ranging from $1000<\ell<3000$, there is an anti-correlation between $n_s$ and $\Omega_bh^2$ for GCMB. 
In addition, we did not find any evidence for the physics beyond the base $\Lambda$CDM cosmology. 


\vspace{5mm}
\noindent {\bf Acknowledgments}

We acknowledge the use of HPC Cluster of Tianhe II in National Supercomputing Center in Guangzhou and HPC Cluster of ITP-CAS. This work is supported by grants from NSFC (grant No. 11975019, 11690021, 11991053, 11947302), the Strategic Priority Research Program of Chinese Academy of Sciences (Grant No. XDB23000000, XDA15020701), and Key Research Program of Frontier Sciences, CAS, Grant NO. ZDBS-LY-7009.



\end{document}